\documentclass[aps,prl,twocolumn,epsfig,amsmath,amssymb]{revtex4}
\usepackage[english]{babel} \usepackage{latexsym}
\usepackage{graphicx} \usepackage{subfigure} \usepackage{epsfig}
\usepackage{psfrag}
\usepackage{subfigure}
 
\begin{document}

\title{Faraday patterns in dipolar Bose-Einstein condensates}
\author{R. Nath and L. Santos} 
\affiliation{
\mbox{Institut f\"ur Theoretische Physik , Leibniz Universit\"at
Hannover, Appelstr. 2, D-30167, Hannover, Germany}\\
}

%
%

\begin{abstract}  
Faraday patterns can be induced in Bose-Einstein condensates by 
a periodic modulation of the system nonlinearity. We show 
that these patterns are remarkably different 
in dipolar gases with a roton-maxon excitation spectrum.
Whereas for non-dipolar gases the pattern size decreases monotonously 
with the driving frequency, patterns in dipolar gases present, 
even for shallow roton minima, a highly non trivial 
frequency dependence characterized by abrupt pattern size transitions, 
which are especially pronounced when 
the dipolar  interaction is modulated.
Faraday patterns constitute hence an optimal tool for revealing 
the onset of the roton minimum, a major key feature of dipolar gases.
\end{abstract}  
\pacs{03.75.Fi,05.30.Jp} 
\maketitle





The formation of Faraday patterns in driven systems 
is a general phenomenon occurring in scenarios ranging from  
hydrodynamics and non-linear optics, to liquid crystals and  
chemical reactions \cite{Cross}. Faraday  
patterns may be observed in Bose-Einstein condensates (BECs)  
by modulating the nonlinearity arising from the interatomic interactions
~\cite{FaradPat} either by 
time-dependent Feshbach resonances~\cite{Feshbach} 
or by a time-dependent confinement.  The latter method 
has been recently used for realizing 
these patterns in BECs~\cite{FaradPatExp}. 
Faraday patterns offer important insights about elementary excitations 
in BECs since the pattern size is determined by 
the Bogoliubov mode resonant with half of the driving frequency. 
For usual short-range interacting BECs the energy of elementary 
excitations grows monotonously as a function of their 
corresponding momenta. As a result, the pattern size 
decreases monotonously with the driving frequency.


Recent experiments on atoms with large magnetic 
moments~\cite{Pfau,ParisNord},
polar molecules~\cite{Jin,Weidemueller}, spinor Bose-Einstein condensates 
(BECs)~\cite{StamperKurn}, and optical lattices~\cite{Fattori}
are opening the rapidly-developing area of dipolar gases. In
these gases, the dipole-dipole interactions (DDIs) play a significant or even
dominant role compared to the short-range interactions (SRIs). Dipolar BECs 
present a wealth of new 
physics~\cite{DipBEC-Yi,DipBEC-Goral,DipBEC-Santos,Roton} due to 
the long-range and anisotropic character of the DDI.
A major salient difference between non-dipolar and dipolar BECs 
is provided by the dispersion of elementary excitations, which,  
due to the momentum dependence of the DDI,  
may show a roton-maxon character~\cite{Roton}, similar to that 
encountered in Helium physics~\cite{Feynman}. 
This key feature of dipolar gases may significantly influence, for 
deep roton minima, the BEC properties at finite 
temperatures~\cite{WangDemler} and even the BEC 
stability~\cite{Komineas}.
The roton minimum has not yet been observed experimentally, and it 
remains still an open question how to probe easily the onset of 
rotonization.


In this Letter, we show that pattern formation is crucially modified  
in dipolar BECs with a roton-maxon spectrum. 
Remarkably, contrary to many pattern forming systems, 
including non-dipolar BECs, 
the first unstable mode does not necessarily determine the emerging pattern, 
which may be dominated by harmonics of the driving frequency with energies 
close to the roton minimum. As a result of that and of the  
multi-valued character of the roton-maxon spectrum 
the pattern size presents a highly non trivial dependence 
with the driving frequency characterized even for shallow roton minima 
by abrupt transitions in the pattern size. 
These transitions, which are especially 
pronounced when the DDI is modulated, may be employed to reveal easily 
the appearance of a roton-minimum in experiments on dipolar BECs.




We consider a BEC of $N$ particles with mass $m$ and 
electric dipole $d$ (the results are equally valid for magnetic dipoles) 
oriented in the $z$-direction by a sufficiently large external field. The 
dipoles interact via a dipole-dipole potential: 
$V_d(\vec{r})= d^2 (1-3\cos^2(\theta))/r^3$, where $\theta$ 
is the angle formed by the vector joining the interacting particles 
and the dipole orientation. We assume a strong harmonic 
confinement $V(z)=m\omega_z^2z^2/2$ 
along the $z$-direction, whereas for simplicity of our discussion 
we consider no $xy$ trapping. At sufficiently low temperatures 
the BEC wavefunction $\Psi(\vec r)$ is given by 
the non-local non-linear Schr\"odinger equation (NLSE):
\begin{equation}
i\hbar\frac{\partial}{\partial t}\Psi(\vec r)=
\left [ \hat H_0 + \int d^3 r' U(\vec r-\vec r') |\Psi(\vec r')|^2
\right ]
\Psi(\vec r),
\label{GPE}
\end{equation}
where $\hat H_0=-\hbar^2\nabla^2/2m+V(z)$ and 
$U(\vec r)=g\delta(\vec r)+ V_d(\vec r-\vec r')$, with 
$g=4\pi\hbar^2a/m$, where $a$ the $s$-wave scattering length 
(we consider in the following $a<0$), and $m$ the particle mass.
If the chemical potential (below
introduced) $\mu_{2d}\ll\hbar\omega_z$, the system can be considered ``frozen'' into the 
ground state $\phi_0(z)$ of $V$ and hence the BEC wave function factorizes 
as $\Psi(\vec r)=\psi(\vec \rho)\phi_{0}(z)$.
Employing this
factorization, the convolution theorem, the Fourier 
transform of the dipole-potential and integrating over 
the $z$ direction, we arrive at the 2D NLSE \cite{Pedri2005}:
\begin{eqnarray}
&&i\hbar\frac{\partial}{\partial t}\psi(\vec \rho)=
\left [ 
-\frac{\hbar^2}{2m}\nabla^2
+g_{2d} |\psi(\vec \rho)|^2 \right\delimiter 0 \nonumber \\
&&+\frac{4\pi}{3} \beta g_{2d}
\left\delimiter 0  \int\frac{d^2 k}{(2\pi)^2}e^{i\vec k \cdot\vec {\rho}} 
\tilde n(\vec k)h_{2d}\left (\frac{k l_z}{\sqrt{2}}\right)
\right ]\psi(\vec \rho),
\label{GPE2D}
\end{eqnarray}
where $\vec k$ is the $xy$-momentum, 
$l_z\equiv\sqrt{\hbar/m\omega_z}$ is the oscillator length,   
$g_{2d}\equiv g/\sqrt{2\pi}l_z$ is 
the 2D short-range coupling constant,  
$\tilde n (\vec k)$ is the Fourier transform of $|\psi(\vec\rho)|^2$, and 
$h_{2d}(\vec k)=2-3\sqrt{\pi}ke^{k^2}$erfc$(k)$, 
with erfc$(k)$ the complementary error function. 
The parameter $\beta=d^2/g$ characterizes the 
DDI strength compared to the SRI.



\begin{figure} 
\begin{center}
\includegraphics[width=6.5cm, height=4cm]{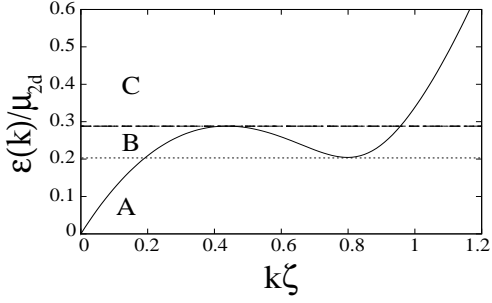}
\end{center} 
\vspace*{-0.2cm} 
\caption{Dispersion of a 2D homogeneous BEC 
of $^{52}$Cr with $a=-0.54$nm ($\beta=-0.375$), 
a 3D density $\bar n_{2d}/\sqrt{2\pi}l_z=10^{14}cm^{-3}$, 
$\hbar\omega_z=\mu_{2d}/2$, and $\zeta=\hbar/\sqrt{2m\mu_{2d}}=0.59 \mu$m.} 
\label{fig:1}
\vspace*{-0.2cm}  
\end{figure}

The homogeneous solution of~(\ref{GPE2D}) is 
$\psi(\vec\rho,t)=\sqrt{\bar n_{2d}}\exp{[-i\mu_{2d} t/\hbar]}$, 
with $n_{2d}$ the 2D density, 
and $\mu_{2d}=g_{2d} n_{2d}(1+8\pi \beta /3)$ the chemical
potential. The elementary excitations of the homogeneous 2D BEC 
are plane waves with 2D wave number $\vec k$ and dispersion~\cite{Fischer}
\begin{equation}
\epsilon(k)^2 = T(k) 
\left [ T(k) +2 g_{2d}n_{2d}
\left [1+\frac{4\pi\beta}{3}h_{2d}\left(\frac{kl_z}{\sqrt 2}\right) \right]
\right ]
\label{DISP1}
\end{equation}
where $T(k)=\hbar^2 k^2/2m$ is the kinetic energy. 
If $\beta=0$ and since $a<0$ then $\epsilon(k\to 0)^2<0$ and 
phonon instability occurs, followed by
the well-known collapse for attractive short-range interacting BECs.
This instability is prevented for sufficiently large DDI 
such that $g+8\pi d^2/3>0$. 
At moderate $d$ values, and 
due to the $k$-dependence of the DDI ($h_{2d}$ function), 
$\epsilon(k)$ may develop a roton-like minimum 
for intermediate $k$ values (see Fig.~\ref{fig:1}).
The roton-maxon spectrum constitutes one of the most relevant novel
features in dipolar gases. We 
show below that this roton minimum may be easily probed 
even for shallow roton minima by modulating 
the system nonlinearity.


We consider a modulation
$a(t)=a_0[1+2\alpha\cos(2\omega t)]$ about its mean 
$a_0$, where $\alpha$ is the modulation amplitude and $2\omega$ 
is the driving frequency.
The homogeneous 2D solution is 
$\psi_{H}(\vec\rho,t)=\sqrt{\bar n_{2d}}
\exp{[-i(t+\frac{\gamma}{\omega}\sin(2\omega t)
)\mu_{2d}/\hbar]}$, with $\gamma=\alpha/(1+8\pi\beta/3)$. 
The driving may induce a dynamical instability breaking 
the translational symmetry. This destabilization 
is best studied with an ansatz
$\psi(\vec \rho,t)=\psi_H(t)[1+w(t)\cos(\vec k\cdot \vec r)]$,
where $w(t)$ is the complex perturbation amplitude. Inserting 
this ansatz into (\ref{GPE2D}) we obtain a Mathieu equation for $u=$Re$(w)$:
\begin{equation}
\frac{d^2u}{dt^2}+\frac{1}{\hbar^2}\left[\epsilon(k)^2+
2 b(\omega,k,\alpha)(\hbar\omega)^2\cos(2\omega t)\right]u=0,
\label{Mat}
\end{equation}
where $b(\omega,k,\alpha)\equiv 2\alpha |g_{2d}|n_{2d}T(k)/(\hbar\omega)^2$, where 
$g_{2d}$ is calculated from the mean $a_0$. 
Following Floquet Theorem, the solutions of (\ref{Mat}) are of the form 
$u(t)=c(t)\exp\sigma t$, where $c(t)=c(t+2\pi/\omega)$ and $\sigma(k,\omega,\alpha)$ 
is the so-called Floquet exponent. If Re$(\sigma)>0$ the homogeneous BEC is 
dynamically unstable against 
the formation of Faraday patterns, whose typical wavelength is dominated by the 
most unstable mode (that with the largest Re$(\sigma)>0$). 
For vanishing modulation the system becomes unstable at the parametric resonances 
$\epsilon(k)=n\hbar\omega$ ($n=1,2,...$).


BECs with repulsive short-range interactions exhibit a spectrum 
$\epsilon(k)^2=T(k)[T(k)+2g_{2d}n_{2d}]$,  characterized 
by phonon-like excitations at low $k$ and single-particle excitations 
at large $k$. For any given driving 
frequency the most unstable mode is always provided by the first
resonance $\epsilon(k)=\hbar\omega$, and hence 
$k=\epsilon^{-1}(\hbar\omega)$  determines the typical 
inverse size of the Faraday pattern~\cite{Nicolin}. As a consequence 
a larger driving frequency leads to a pattern of smaller size, 
as shown in recent experiments~\cite{FaradPat}.
In the following we show that the physics of Faraday patterns is 
remarkably much richer and involved in dipolar BECs.

\begin{figure} 
\begin{center}
\subfigure{
\includegraphics[width=4.cm, height=3.6cm]{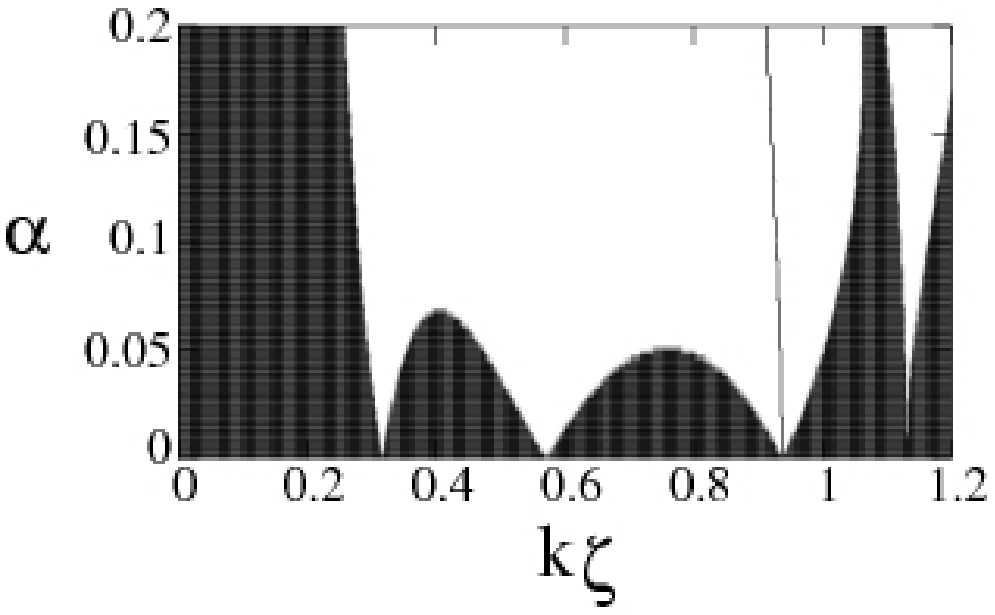}
\label{fig2:subfig1}}
\hspace{-0.1cm}
\subfigure{
\includegraphics[width=4.cm, height=3.6cm]{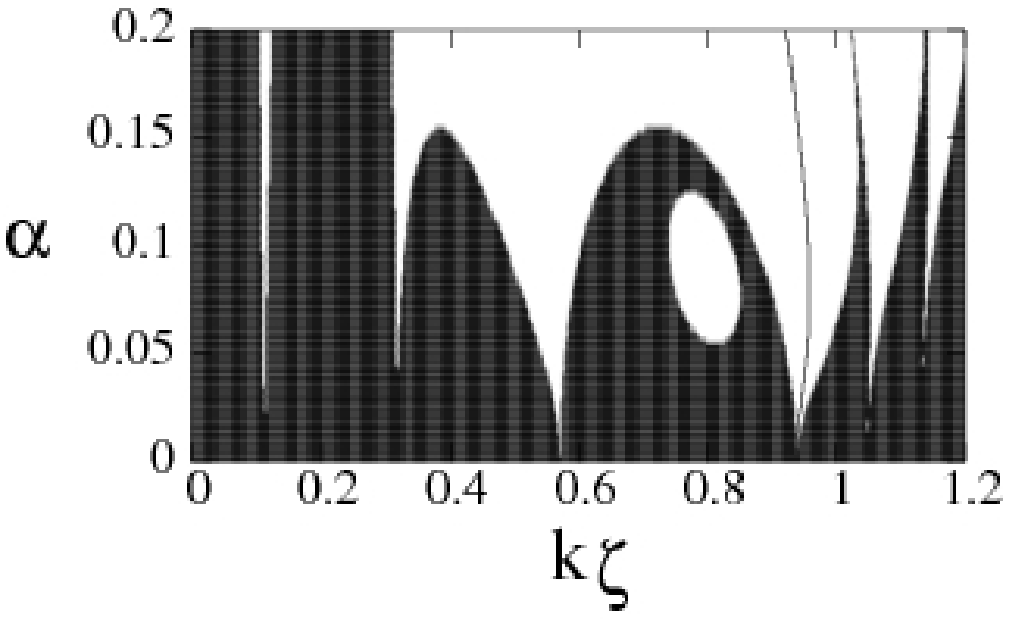}
\label{fig2:subfig2}}
\end{center} 
\vspace*{-0.2cm} 
\caption{Stability diagram (dark regions are stable)
for the parameters of Fig.~\ref{fig:1} as a function of $\alpha$ and $k$.   
(Left) $\hbar\omega/\mu=0.268$ (regime B). 
(Right) $\hbar\omega/\mu=0.134$ (regime A). The most unstable 
modes are indicated by a solid line.} 
\label{fig:2}
\vspace*{-0.2cm}  
\end{figure}

As mentioned above, for intermediate $d$ values $\epsilon(k)$ 
shows a roton minimum (with energy $\hbar\omega_r$) and a maxon maximum ($\hbar\omega_m$). 
Hence, as a function of the modulation frequency $2\omega$ we may distinguish 
three driving regimes: (A) $\omega<\omega_r$, 
(B) $\omega_r\leq\omega\leq\omega_m$ and (C) $\omega>\omega_m$ (see Fig.~\ref{fig:1}). 
The latter regime is relatively uninteresting, 
since, as for non-dipolar BECs, the spectrum is uni-valued and 
the most unstable mode is provided by $\epsilon(k)=\hbar\omega$.
The regime B on the contrary is multi-valued, and the condition 
$\epsilon(k)=\hbar\omega$ is satisfied by a triplet 
$k_1<k_2<k_3$. These three resonant momenta lead to three instability tongues 
for growing modulation amplitude $\alpha$ (Fig.~\ref{fig2:subfig1}). 
Hill's solution method~\cite{Nicolin,NicolinThesis} provides that for the lowest resonance 
$\epsilon(k)=\hbar\omega$, the Floquet exponent may be approximated for small 
$b$ as $\sigma_1=b(\omega,k,\alpha)/2\propto k^2/(\hbar\omega)^2$. Hence 
the most unstable mode in regime B is always given by the largest momentum 
$k_3$, which dominates the Faraday pattern formation.

\begin{figure} 
\begin{center}
\subfigure{
\includegraphics[width=4.cm, height=3.3cm]{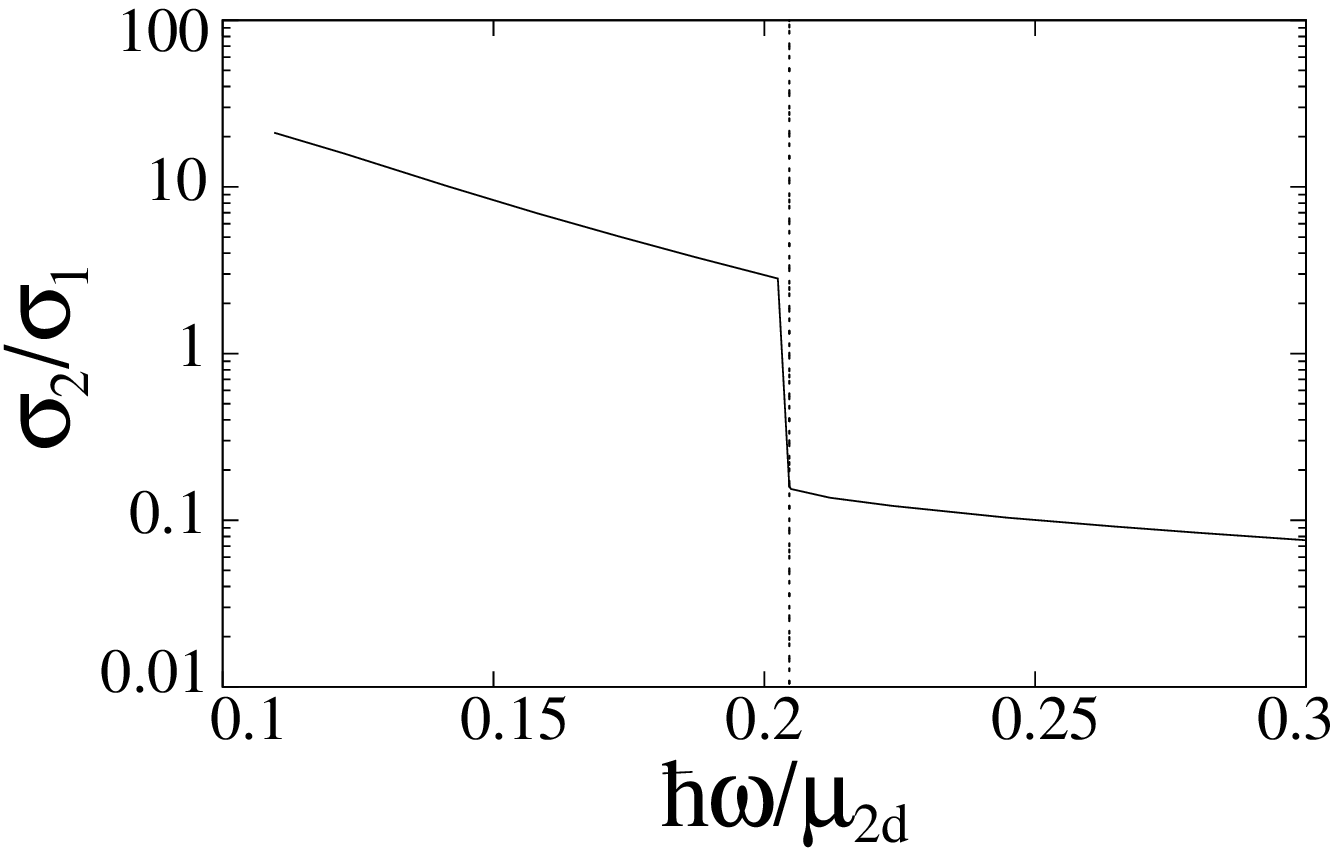}
\label{fig3:subfig1}}
\hspace{-0.1cm}
\subfigure{
\includegraphics[width=4.cm, height=3.3cm]{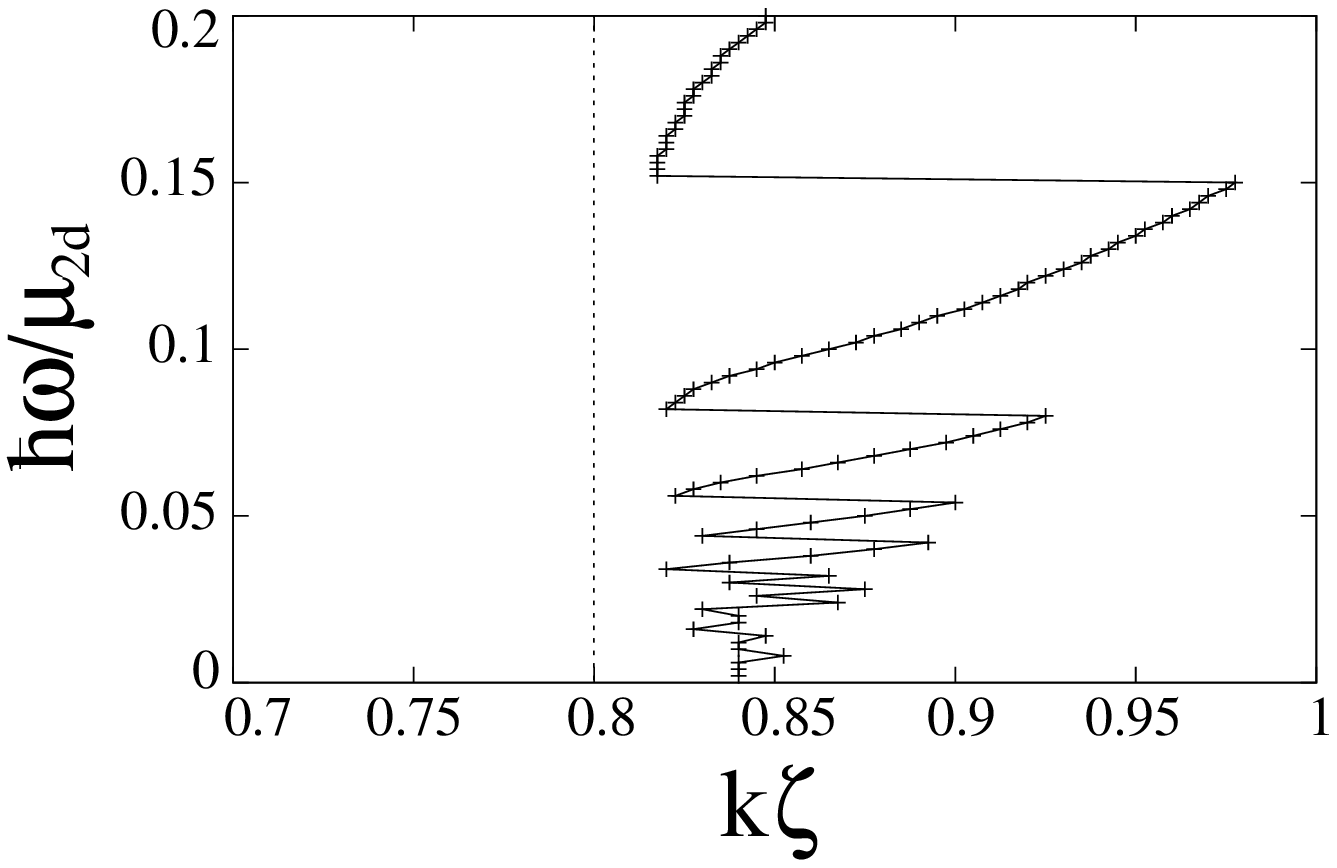}
\label{fig3:subfig2}}
\end{center} 
\vspace*{-0.2cm} 
\caption{(Left) Ratio between the Floquet exponent $\sigma_1$ 
for $\epsilon=\hbar\omega$ 
and $\sigma_2$ for $\epsilon=2\hbar\omega$. 
(Right) Most unstable $k$ as a function of $\omega$. We 
use the same parameters as for Fig.~\ref{fig:1}. The dashed line 
indicates the roton frequency or momentum.} 
\label{fig:3}
\vspace*{-0.2cm}  
\end{figure}

For regimes B and C the Faraday pattern is, 
as for non-dipolar BECs, provided 
by $\epsilon(k)=\hbar\omega$. The situation is remarkably different 
for regime A. The latter is better understood by considering the 
ratio $\sigma_2/\sigma_1$ between the Floquet exponents for 
the second ($\epsilon(k)=2\hbar\omega$) and the first 
($\epsilon(k)=\hbar\omega$)
resonance condition. This ratio may be obtained by using again 
Hill's solution method~\cite{Nicolin,NicolinThesis}:
\begin{equation}
\frac{\sigma_2}{\sigma_1}=
\frac{\sqrt{5}\alpha}{12(8\pi|\beta|/3-1)}\left (\frac{\mu_{2d}}{\hbar\omega}\right )^2
\zeta^2 \frac{[\epsilon^{-1}(2\hbar\omega)]^4}{[\epsilon^{-1}(\hbar\omega)]^2}
\end{equation}
where $\zeta=\hbar/\sqrt{2m\mu_{2d}}$ is the healing length. 
Not surprisingly, the first resonance dominates for $\alpha\rightarrow 0$. 
However, contrary to the short-range interacting case, 
for $\alpha$ surpassing a very small $\omega$-dependent critical $\alpha$ 
the situation changes completely. 
Fig.~\ref{fig3:subfig1} depicts the ratio $\sigma_2/\sigma_1$ as a function of 
$\omega$ for a small $\alpha=0.04$. Note that for $\omega>\omega_r$ 
$\sigma_2<\sigma_1$, and as expected, for regimes B and C 
the instability is dominated by 
the resonance $\epsilon(k)=\hbar\omega$. On the contrary for 
$\omega<\omega_r$, $\sigma_2>\sigma_1$ even for such a small value 
of $\alpha$, and hence the lowest resonance is not any more the 
most unstable one. This surprising result 
is a direct consequence of the non-monotonous character of the 
roton-maxon dispersion law.

Our numerical Floquet analysis shows indeed 
(see Fig.~\ref{fig3:subfig2}) that 
for $\alpha>\alpha_{cr}$ (for the parameters of Figs.~\ref{fig:2} $\alpha_{cr}\simeq 0.027$)  
the most unstable mode for all driving frequencies within the regime A is given by the 
largest momenta $k$ compatible with the first harmonic 
$\epsilon(k)=n\hbar\omega$ lying in the regime B (or, if none, 
the first lying in regime C). This has important consequences for the 
wavenumber selection as a function of the driving $\omega$, which, 
as shown in Fig.~\ref{fig3:subfig2}, is remarkably different than that for the 
non-dipolar case. The pattern size does not decrease monotonously 
with growing $\omega$, but on the contrary oscillates 
in the vicinity of the roton momentum, presenting abrupt changes of the pattern size 
at specific driving frequencies. These oscillations are the result of 
the subsequent destabilization of higher harmonics in regime B.



\begin{figure} 
\begin{center}
\subfigure{
\includegraphics[width=4.cm, height=3.8cm]{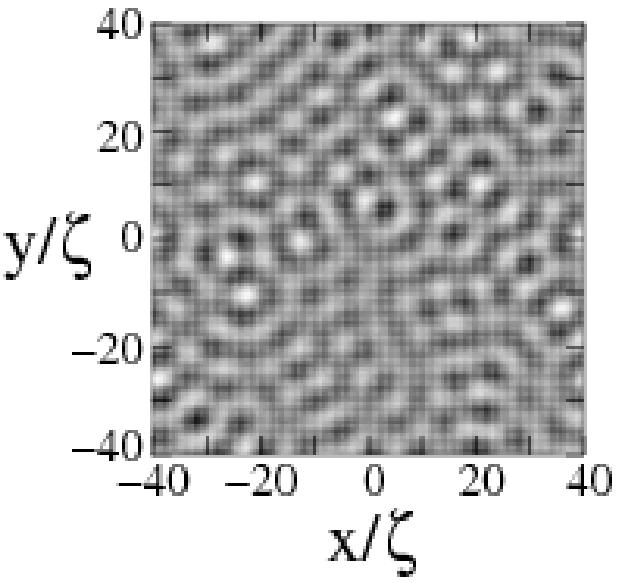}
\label{fig4:subfig1}}
\hspace{-0.1cm}
\subfigure{
\includegraphics[width=4.cm, height=3.8cm]{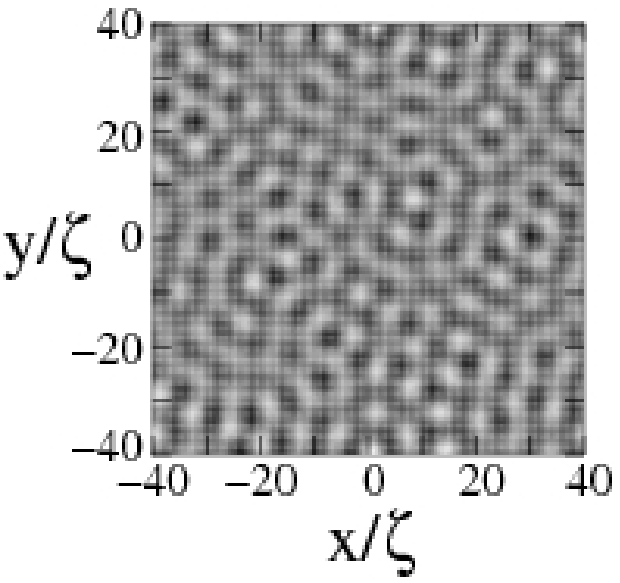}
\label{fig4:subfig2}}
\end{center} 
\vspace*{-0.2cm} 
\caption{Faraday patterns 
for $\hbar\omega/\mu=0.268$ at $t=40.6$ ms (left) 
and $\hbar\omega/\mu=0.134$ at $t=93.2$ ms (right), 
and the same 
parameters as in Fig.~\ref{fig:1}.}
\label{fig:4}
\vspace*{-0.2cm}  
\end{figure}

This distorted wave number selection is directly mirrored into the 
spatial form of the corresponding Faraday patterns. 
We have studied the dynamical instability induced by the modulation and 
the corresponding Faraday patterns by 
simulating Eq.~(\ref{GPE2D}) numerically with 
periodic boundary conditions and an overimposed random noise provided 
by a tiny random local phase ($<10^{-3}\pi$) on the homogeneous solution. 
Our direct numerical calculations is in excellent agreement with our 
Floquet analysis. Fig.~\ref{fig4:subfig1} depicts the case of 
a frequency $\omega_r<\omega_0<\omega_m$, where as expected 
the Faraday pattern is indeed given by the largest resonant momentum.  
In Fig.~\ref{fig4:subfig2} 
we consider $\omega=\omega_0/2$ which is within the regime A. 
Strikingly, due to the discussed selection of higher harmonics,
the Faraday pattern is basically the same as for a double driving frequency $\omega_0$.
This quasi-insensitivity becomes quantitatively evident after Fourier transforming the pattern.

In the previous discussion we considered the modulation 
of the $s$-wave scattering length $a(t)$. A dipolar BEC offers, however, 
an additional novel way of modifying the system nonlinearity by a time-dependent 
DDI. This may be achieved by modulating slightly the intensity of the 
polarizing field (e.g. the electric field orienting a polar molecule) 
or by introducing a 
slight precession of the direction of the external field (e.g. by 
additional transversal magnetic fields in the case of atomic dipoles). 
In the following we show that the Faraday patterns obtained by means 
of a modulated DDI differ very significantly from those obtained by 
modulating $a(t)$.

\begin{figure} 
\begin{center}
\includegraphics[width=5.5cm, height=3.5cm]{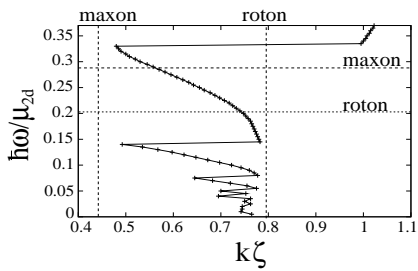}
\end{center} 
\vspace*{-0.2cm} 
\caption{Most unstable $k$ as a function of $\omega$ for modulated 
$\beta(t)$, $\alpha=0.12$, and the same parameters as for 
Fig.~\ref{fig:1}. We indicate the roton and 
maxon frequencies and momenta.}
\label{fig:5}
\vspace*{-0.2cm}  
\end{figure}

\begin{figure} 
\begin{center}
\subfigure{
\includegraphics[width=4.0 cm, height=3.7cm]{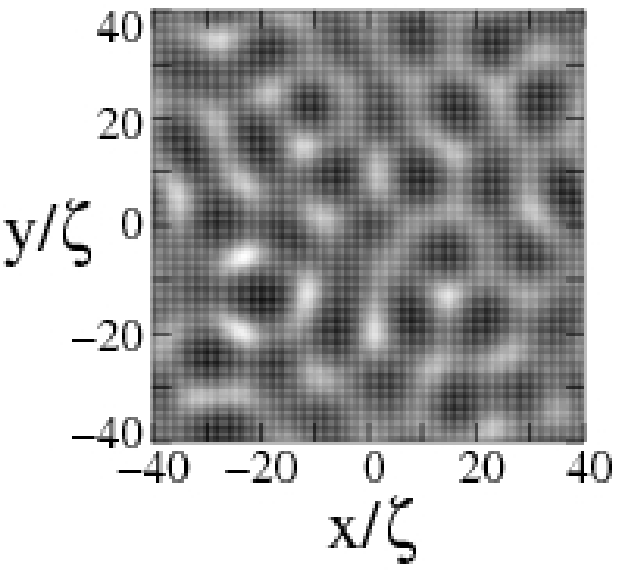}
\label{fig6:subfig1}}
\hspace{-0.1cm}
\subfigure{
\includegraphics[width=4.0 cm, height=3.8cm]{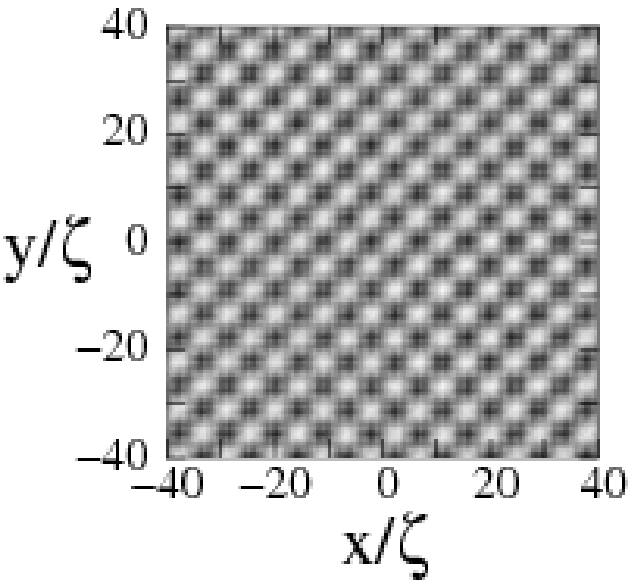}
\label{fig6:subfig2}}
\end{center} 
\vspace*{-0.2cm} 
\caption{Faraday patterns 
for $\hbar\omega/\mu=0.32$ at $t=116.7$ ms (left) 
and $\hbar\omega/\mu=0.34$ at $t=405.1$ ms (right), 
and the same parameters as in Fig.~\ref{fig:5}.}
\label{fig:6}
\vspace*{-0.2cm}  
\end{figure}

We consider a temporal modulation of the DDI $d^2=g\beta(t)$, with 
$\beta(t)=\bar{\beta}[1+2\alpha\cos(2\omega t)]$ about its 
mean value $\bar{\beta}$. Following a similar procedure as that 
discussed above for the case of modulated $a(t)$, we obtain the 
Mathieu equation for the real part of the perturbation amplitude. 
This equation is of the same form as Eq.~(\ref{Mat}) but 
with 
\begin{equation}
b(\omega,k,\alpha)=\frac{8\pi\alpha |\bar{\beta}g_{2d}|n_{2d}}{3(\hbar\omega)^2}
T(k)h_{2d} \left ( \frac{kl_z}{\sqrt{2}} \right ).
\end{equation}
The modified $k$-dependence of $b(\omega,k,\alpha)$ has 
crucial consequences 
for the formation of Faraday patterns. 
Similarly as above we may obtain from Hill's solution method 
the Floquet exponent for the first resonance 
$\epsilon(k)=\hbar\omega$, $\sigma_1\propto k^2 h_{2d}(kl_z/\sqrt{2})$. 
This leads to a remarkably different selection rule for 
$\omega$ values within the regime B. Contrary to the 
case of modulated $a(t)$, it is the intermediate momentum $k_2$ and not the 
largest one $k_3$ the most unstable within regime B. This leads to a 
remarkably abrupt change in the Faraday pattern size in the vicinity of $\omega_m$~\cite{footnote}.  
In addition, and similar to the case of modulated $a(t)$, driving with 
$\omega<\omega_r$ may be dominated by higher harmonics. 
Fig.~\ref{fig:5} shows the most unstable $k$ 
as a function of $\omega$ for a typical case of modulated $\beta(t)$. Note 
not only the above mentioned abrupt jump in the vicinity of $\omega_m$ but also 
at other $\omega$ values within the regime A. As for the case of 
modulated $a(t)$ these jumps represent abrupt transitions 
in the Faraday pattern size, which are 
certainly much more marked than for the modulated $a$ case. 
Figs.~\ref{fig:6} show the abrupt change in the patterns for two driving 
frequencies right below and above the transition close to 
$\omega_m$~\cite{footnote}.

In our calculations we have assumed for simplicity no trapping on the 
$xy$ plane. 
An harmonic $xy$-confinement (with frequency $\omega_\perp$) 
leads to a finite momentum cut-off $k_{c}=\sqrt{m\omega_\perp/\hbar}$. 
In a good approximation all features in the excitation spectrum with momenta 
$k\gg k_{c}$ are not affected by the inhomogeneous trapping. 
For typical roton momenta $k\zeta\simeq 0.5$ and 
$\zeta\simeq 0.6\mu m$ in our figures, $k\gg k_c$ demands for $^{52}$Cr 
a transversal frequency $\omega_\perp < 130$Hz, which can be considered 
a typical experimental condition.
Finally, we stress that 
Faraday patterns are a transient phenomenon, and that for the case  
discussed here ($a<0$) pattern formation is followed 
by collapse (and consequent violation of the two-dimensional condition).


In summary, the physics of Faraday patterns is largely modified 
in dipolar BECs in the presence of even shallow roton minima. Whereas 
in non-dipolar BECs the 
Faraday pattern size decreases monotonously with the 
driving frequency $2\omega$, 
in dipolar BECs the patterns show a $\omega$-dependence
characterized by abrupt changes in the pattern size, which 
are especially remarkable when the dipole itself is modulated. Faraday patterns 
constitute hence an excellent tool to probe the onset of 
rotonization in on-going experiments with dipolar condensates.


\acknowledgements

This work was supported by the DFG (SFB407, QUEST), and the ESF (EUROQUASAR).


\end{document}